\documentclass[letterpaper,preprint,prl,aps,superscriptaddress,floatfix]{revtex4}
\usepackage[latin1]{inputenc}
\usepackage{bm}
\usepackage{multirow,amssymb,amsbsy,amsmath}
\usepackage{graphicx}
\usepackage{verbatim}
\usepackage{color}

\makeatletter
\usepackage{pifont}
\makeatother

\usepackage{graphicx}
\usepackage{dcolumn}
\usepackage{bm}
\usepackage{ifthen}
\usepackage{booktabs}
\usepackage{nicefrac}
\usepackage{float}
\usepackage{bbm}
\usepackage{amsmath}
\usepackage{ulem}

\definecolor{gray}{rgb}{0.75,0.75,0.75}

\begin{document}

\title{Certification of Genuine Multipartite Entanglement with General and Robust Device-independent Witnesses}
\author{Chao Zhang}
\thanks{These authors contributed equally to this work}
\affiliation{CAS Key Laboratory of Quantum Information, University of Science and Technology of China, Hefei, Anhui 230026, China.}
\affiliation{CAS Center For Excellence in Quantum Information and Quantum Physics, University of Science and Technology of China, Hefei, Anhui 230026, China.}
\author{Wen-Hao Zhang}
\thanks{These authors contributed equally to this work}
\affiliation{CAS Key Laboratory of Quantum Information, University of Science and Technology of China, Hefei, Anhui 230026, China.}
\affiliation{CAS Center For Excellence in Quantum Information and Quantum Physics, University of Science and Technology of China, Hefei, Anhui 230026, China.}
\author{Pavel Sekatski}
\email{pavel.sekatski@gmail.com}
\affiliation{Department of Applied Physics, University of Geneva,
Rue de l'\'Ecole-de-M\'edecine, 1211 Geneva, Switzerland}
\author{Jean-Daniel Bancal}
\email{jdbancal.physics@gmail.com}
\affiliation{Universit\'e Paris-Saclay, CEA, CNRS, Institut de physique th\'eorique, 91191, Gif-sur-Yvette, France}
\author{Michael Zwerger}
\email{michael.zwerger@mpl.mpg.de}
\affiliation{Max Planck Institute for the Science of Light, Erlangen, Germany}
\author{Peng Yin}
\affiliation{CAS Key Laboratory of Quantum Information, University of Science and Technology of China, Hefei, Anhui 230026, China.}
\affiliation{CAS Center For Excellence in Quantum Information and Quantum Physics, University of Science and Technology of China, Hefei, Anhui 230026, China.}
\author{Gong-Chu Li}
\affiliation{CAS Key Laboratory of Quantum Information, University of Science and Technology of China, Hefei, Anhui 230026, China.}
\affiliation{CAS Center For Excellence in Quantum Information and Quantum Physics, University of Science and Technology of China, Hefei, Anhui 230026, China.}
\author{Xing-Xiang Peng}
\affiliation{CAS Key Laboratory of Quantum Information, University of Science and Technology of China, Hefei, Anhui 230026, China.}
\affiliation{CAS Center For Excellence in Quantum Information and Quantum Physics, University of Science and Technology of China, Hefei, Anhui 230026, China.}
\author{Lei Chen}
\affiliation{CAS Key Laboratory of Quantum Information, University of Science and Technology of China, Hefei, Anhui 230026, China.}
\affiliation{CAS Center For Excellence in Quantum Information and Quantum Physics, University of Science and Technology of China, Hefei, Anhui 230026, China.}

\author{Yong-Jian Han}
\affiliation{CAS Key Laboratory of Quantum Information, University of Science and Technology of China, Hefei, Anhui 230026, China.}
\affiliation{CAS Center For Excellence in Quantum Information and Quantum Physics, University of Science and Technology of China, Hefei, Anhui 230026, China.}
\author{Jin-Shi Xu}
\affiliation{CAS Key Laboratory of Quantum Information, University of Science and Technology of China, Hefei, Anhui 230026, China.}
\affiliation{CAS Center For Excellence in Quantum Information and Quantum Physics, University of Science and Technology of China, Hefei, Anhui 230026, China.}
\author{Yun-Feng Huang}
\email{hyf@ustc.edu.cn}
\affiliation{CAS Key Laboratory of Quantum Information, University of Science and Technology of China, Hefei, Anhui 230026, China.}
\affiliation{CAS Center For Excellence in Quantum Information and Quantum Physics, University of Science and Technology of China, Hefei, Anhui 230026, China.}
\author{Geng Chen}
\email{chengeng@ustc.edu.cn}
\affiliation{CAS Key Laboratory of Quantum Information, University of Science and Technology of China, Hefei, Anhui 230026, China.}
\affiliation{CAS Center For Excellence in Quantum Information and Quantum Physics, University of Science and Technology of China, Hefei, Anhui 230026, China.}
\author{Chuan-Feng Li}
\email{cfli@ustc.edu.cn}
\affiliation{CAS Key Laboratory of Quantum Information, University of Science and Technology of China, Hefei, Anhui 230026, China.}
\affiliation{CAS Center For Excellence in Quantum Information and Quantum Physics, University of Science and Technology of China, Hefei, Anhui 230026, China.}
\author{Guang-Can Guo}
\affiliation{CAS Key Laboratory of Quantum Information, University of Science and Technology of China, Hefei, Anhui 230026, China.}
\affiliation{CAS Center For Excellence in Quantum Information and Quantum Physics, University of Science and Technology of China, Hefei, Anhui 230026, China.}

\date{\today}

\begin{abstract}
Genuine multipartite entanglement represents the strongest type of entanglement, which is an essential resource for quantum information processing. Standard methods to detect genuine multipartite entanglement, e.g., entanglement witnesses, state tomography, or quantum state verification, require full knowledge of the Hilbert space dimension and precise calibration of measurement devices, which are usually difficult to acquire in an experiment.
The most radical way to overcome these problems is to detect entanglement solely based on the Bell-like correlations of measurement outcomes collected in the experiment, namely, device-independently (DI). However, it is difficult to certify genuine entanglement of practical multipartite states in this way, and even more difficult to quantify it, due to the difficulty to identify optimal multipartite Bell inequalities and protocols tolerant to state impurity. In this work, we explore a general and robust DI method which can be applied to various realistic multipartite quantum state in arbitrary finite dimension, while merely relying on bipartite Bell inequalities. Our method allows us both to certify the presence of genuine multipartite entanglement and to quantify it.
Several important classes of entangled states are tested with this method, leading to the detection of genuinely entangled states. We also certify genuine multipartite entanglement in weakly-entangled GHZ states, thus showing that the method applies equally well to less standard states.
\end{abstract}

\maketitle

\section{Introduction}
Genuine multipartite entanglement (GME) is a topic of intense research because of its potential impact in quantum computation and condensed matter physics. Currently available techniques have realized Schr\"odinger cat states of up to 20 qubits \cite{Song,Omran}. A natural question arising in such experiments is how to certify the presence of GME . Usual solutions  consist in  measuring a witness \cite{Guhne,Toth1,ZYY} of GME, doing a full state tomography followed by further analysis of the reconstructed density matrix \cite{Hradil,Paris,Hou}, or executing quantum state verification on the premise of accessing some partial prior knowledge   about the state \cite{ZWH,Hayashi,Pallister,ZHJ}. However, these approaches require sufficient knowledge of both the internal physical structure of the measurement devices, and the dimension for the Hilbert space of each system \cite{Moroder}. Unfortunately, it is usually difficult to access an exact quantum description of measurement devices, since the actual measurement settings may deviate from the expected ones slightly and result in an incorrect conclusion about the tested states~\cite{Rosset12}. Furthermore, a physical system typically has access to more levels and degrees of freedom than one uses to describe its state, e.g., a photon has many degrees of freedom (polarization, position, orbital angular momentum, energy levels, etc.), therefore it is questionable to simply view each photon as a qubit. In fact, using an inappropriate description of the system at hand can have devastating consequence when using the system for quantum applications, as demonstrated in recent hacking experiments~\cite{Lydersen2010,Gerhardt2011}.

In order to circumvent this problem, researchers opened a new realm of quantum science, namely ``device-independent" science \cite{Brunner14,Acin1,Acin2,Masanes,Pironio,Lunghi1,Pal,Chen,Rabelo,Rabelo1}, in which no assumptions are made about the states under observation, the experimental measurement devices, or even the dimensionality of the Hilbert spaces where such elements are defined. In this approach, the only way to study a system is to perform local measurements on well-separated subsystems and analyze the statistical results. Many theoretical and experimental efforts have been devoted to device-independent certification (DIC) of bipartite entangled states based on Bell tests  \cite{Gisin1991,Gisin1992,Bell,Popescu1,Brunner1}. But it is still a formidable challenge to extend this method to general multipartite scenarios. The  difficulty mainly results from the lack of multipartite Bell inequalities tailored to arbitrary quantum states. Furthermore, for the known inequalities the complexity (number of different measurement to perform) typically increases exponentially with the number of parties, making them impractical. To date, DIC has been intensively investigated for a few simple types  of multipartite entangled states \cite{Gachechiladze,Popescu,Choudhary,Li}, and several specific genuinely entangled states \cite{Bancal,Bancal2011PRL,Liang,Luo} have been investigated. Similarly, self-testing, an approach allowing one to identify the quantum state device-independently was only pursued for a few states \cite{Supic20,Jed,Coladangelo,ZWH1,ZWH2,Baccari20}. Recently, a dissociated DIC (DDIC) method to detect GME for pure states was proposed whereby the detection of GME is reduced into a set of bipartite problems, for each of which a bipartite Bell inequality is tested \cite{Zwerger}. This scheme applies to all multipartite pure states in arbitrary finite dimension and tolerates non-maximal violation; however, the level of admissible noise is limited.

In this work, we build on the technique of~\cite{Zwerger} and propose a generalized DDIC method. Crucially, our method enhances the robustness to noise while still allowing for the detection of arbitrary pure entangled states.  Then, using the polarization of single photons, we test several essential entangled states experimentally, including GHZ states, partially entangled states and cluster states.

The limited detection efficiency in our setup leads to the occurrence of no-click events in the experiment. Simply rejecting these events (post-selection) opens the infamous detection loophole, and forbids the DI analysis of the post-selected measurement data. This problem is overcome here by introducing a minimal assumption on the internal functioning of the measurement devices, the so-called weak fair-sampling assumption~\cite{Orsucci}, which is well justified in our setup, see methods.

\section{Results}

\subsection{Device independent certification of GME}

The main target of this work is to distinguish genuinely entangled states from biseparable ones, which can be expressed as
\begin{equation}
\label{bisep}
\rho_{BS}=\sum_{g_1,g_2} P_{g_1|g_2} \sum_\lambda P(\lambda) \rho_{g_1}(\lambda) \otimes \rho_{g_2}(\lambda),
\end{equation}
where the groups $g_1\cup g_2 =\{1,\dots, N\}$ form a bipartition of the $N$ parties, the first sum runs over all such splittings,  $\rho_{g_1}$ and $\rho_{g_2}$ are arbitrary quantum states of the parties belonging to the respective group, and $\lambda$ is a variable distributed accordingly to $P(\lambda)$. By definition, genuinely multipartite entangled states cannot be written in this form, and involve a contribution that does not split as a tensor product for any bipartition. To illustrate this phenomenon, consider the case $N=4$ with  the general decomposition
\begin{equation}
\begin{split}
\label{mix}
&\rho= P_{ABCD} \, \rho_{ABCD} + P_{AB|CD} \, \rho_{AB|CD} + P_{AC|BD}\,  \rho_{AC|BD}
+ P_{AD|BC}\,  \rho_{AD|BC} \\
& + P_{ABC|D}\,  \rho_{ABC|D} + P_{ABD|C }\, \rho_{ABD|C}
+ P_{ACD|B}\,  \rho_{ACD|B} + P_{A|BCD}\,  \rho_{A|CBD}. \\
\end{split}
\end{equation}
It is convenient to represent the terms in the decomposition with a graph $G$ as depicted in Fig. 1(a).
The state is genuinely multipartite entangled iff the lowest possible value of $P_{ABCD}$ in the decomposition is larger than zero.We thus now only consider decompositions where $P_{ABCD}$ is minimal.

A genuinely multipartite entangled state (GME state) can be device-independently certified as we will now show. To start, we chose a covering $E$ -- a set of pairs of parties (edges) defining  a graph connecting all parties. Then, we aim to  reveal bipartite Bell nonlocality for each edge $e =\{i,j\} \in E$. To achieve this, the remaining parties in $R \subset \{1,...,N\}\text{\textbackslash} e$ are first measured in order to leave the parties $e$ in a pure entangled state~\footnote{In general, a sequence of local operation with classical communication (LOCC) may be performed on the remaining particles $R$}.
 For each branch, defined by the combination of measurement outcomes on $R$ , we test some bipartite Bell inequality between the parties in $e$ with fixed local bound $\beta^L$ and quantum bound $\beta^Q > \beta^L$.  The Bell score $\beta_e$ associated to the edge $e$, is then defined as the average of the Bell scores obtained over all branches. In the ideal case all the Bell tests can be chosen such that $\beta_e=\beta^Q$. Finally, the observation of a large enough average score $\bar{\beta}^E = \frac{1}{|E|} \sum_{e\in E} \beta_e$ over all pairs $e\in E$ allows one to infer that the underlying state is GME~\cite{Zwerger}.

Indeed, if the measured state can be decomposed in the form of Eq.~\eqref{bisep} (with $P_{ABCD}=0$ in the example), each term  $\rho_{g_1}\otimes \rho_{g_2}$ in the decomposition  "cuts" at least one edge $e \in E$. More precisely, there is at least one pair $e =\{i,j\}$ with the two parties belonging to different groups $i \in g_1$ and $j \in g_2$. For this term the Bell score $\beta_e$ can never exceed the local bound $\beta^L<\beta^Q$. As this happens for each term, the biseparable bound is necessarily lower than the quantum maximum $\bar{\beta}_{BS}^E< \beta^Q$, and observing a value $\bar{\beta}^E$ exceeding $\bar{\beta}_{BS}^E$ proves GME.
However, the precise value of this biseparable bound depends on the chosen covering $E$.

To see this, consider two extreme cases of coverings: the minimal covering $E_\text{mini}$ with $|E_\text{mini}|=N-1$ edges (minimally connected graph) and the full covering $E_\text{full}$ with $|E_\text{full}|=N(N-1)/2$ edges (fully connected graph). In the case of minimal covering, one can always find a bipartition that only cuts one edge, see Fig. 1(b). Therefore for $E_\text{mini}$ the biseparable bound on the average Bell score is given by
\begin{equation}
\label{mini}
\bar{\beta}_{BS}^{mini}=\frac{\left(|E_\text{mini}|-1\right) \beta^Q + \beta^L}{|E_\text{mini}|} = \beta^Q -\frac{\beta^Q-\beta^L}{N-1}
\end{equation}
 In the case of a full covering any bipartition cuts at least $N-1$ edges, see Fig.1 (c). Normalizing by the number of pairs the biseparable bound for a full covering reads
\begin{equation}
\label{full}
\bar{\beta}_{BS}^{full}=\frac{\left(|E_\text{full}|-(N-1)\right) \beta^Q + (N-1)\beta^L}{|E_\text{full}|} = \beta^Q -2\,  \frac{\beta^Q-\beta^L}{N}
\end{equation}
It can be easily seen that $\bar{\beta}_{BS}^{full}$ is lower than $\bar{\beta}_{BS}^{mini}$, which is a natural result since measuring more edges  reveals more information about the structure of  the state, which helps  to certify GME. Hence, the full covering is more tolerant to noise. In the Appendix we show that the biseparable bound for the full covering is optimal, as one naturally expects. I.e. there is no covering $E$ for which a lower average violation $\bar{\beta}^E \leq \bar{\beta}^{full}_{BS}$ certifies GME.

Note that these required violations $\bar{\beta}_{BS}^{full}$ and $\bar{\beta}_{BS}^{mini}$ are strictly larger than the local bound $\beta^L$, as one would expect. Indeed, biseparable states can be mixtures of states which are separable according to different partitions, and can therefore produce some Bell violation for each pairs of parties.
For instance, consider the tripartite state
\begin{equation}
\begin{split}
\label{mixed}
\rho_{BS}=
\frac{1}{3}\Big(&|\Phi^+\rangle\langle \Phi^+|_{AB}\otimes|0\rangle \langle 0|_{C} \otimes |1\rangle \langle 1 |_{A'B'C'}  +
|\Phi^+\rangle\langle \Phi^+|_{AC}\otimes|0\rangle \langle 0|_{B} \otimes |1\rangle \langle 1 |_{A'B'C'} \\
+ &|\Phi^+\rangle\langle \Phi^+|_{BC}\otimes|0\rangle \langle 0|_{A} \otimes |2\rangle \langle 2 |_{A'B'C'}  \Big)
\end{split}
\end{equation}
in which $|\Phi^+\rangle$ is a maximally entangle two-qubit states, and the qutrit "label" systems are prepared in product states $|i\rangle_{A'B'C'} = |i\rangle_{A'} |i\rangle_{B'} |i\rangle_{C'} $. This state is manifestly biseparable since it is a mixture of three biseparable components. Yet, for each edge in a full covering (AA'-BB', AA'-CC',  BB'-CC') the violation can be as high as $\beta_e = \frac{1}{3} \beta^Q + \frac{2}{3} \beta^L >\beta^L$. In fact, this bound saturates the biseparable bound $\bar{\beta}^{full}_{BS}$. When the measurements of the two parties performing the Bell tests are allowed to depend on the preparation branch it can be checked that the biseparable bounds $\bar\beta_{BS}^{mini}$ and $\bar\beta_{BS}^{full}$ are tight for any $N$ by engineering biseparable states in this manner.

Whereas the full covering comes with a higher tolerance against imperfections it also makes the protocol less practical. Indeed, the number of pairs in the covering set increases from a number linear in $N$ to a quadratic number. Clearly, the minimal and full covering are the two limiting cases of a general covering set in this respect. Interestingly, we show in the Appendix that a ring covering $E_{ring}$ also saturates the optimal biseparable bound $\bar{\beta}^{ring}_{BS}=\bar{\beta}^{full}_{BS}$, while only involving a linear number of edges $|E_{ring}|=N$. The  number of parties that have to be measured in order to prepare an entangled state between some pair $e$ appearing in the covering also scales with the system size in general.  For practical purposes it is helpful if entangled states can be prepared by operating on a small number of parties for all pairs added to the covering. This can be done generically for interesting states, such as generalized, weighted graph states with bounded degree.

It is worth noting, however, that all states preparations need not be evaluated in practice in order to conclude about GME: since every such preparation across a biseparation must satisfy the local bound, finding one violation is enough to conclude about GME. In other words, a statistically significant violation of our witness over one edge $e$ can be concluded after a number of samplings that is fixed, i.e. that doesn't scale with $N$.

\begin{figure}[htbp]
\centering
\includegraphics[width=5in]{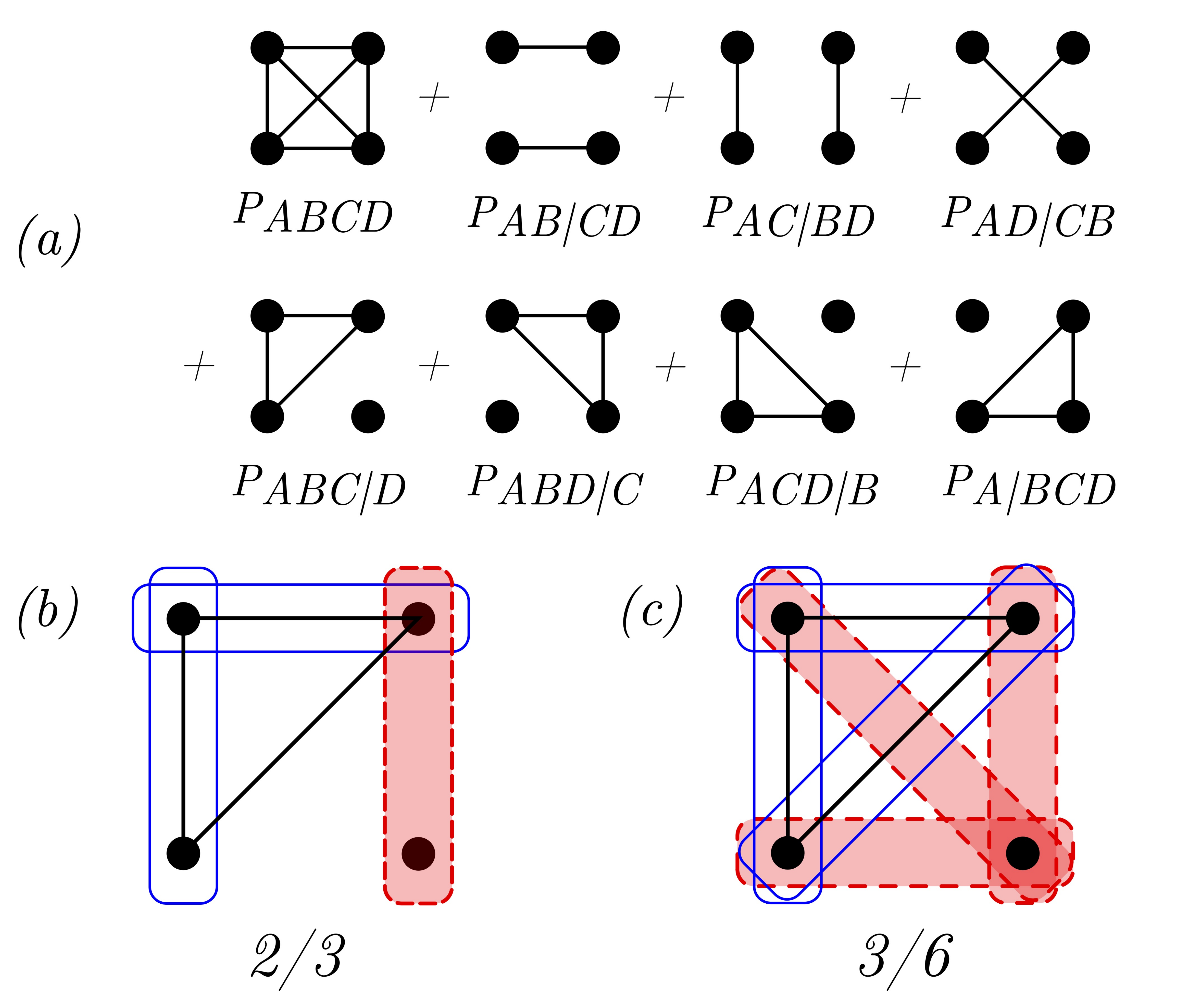}
\caption{\textbf{Graph representation of multipartite states.}  (a) Graphical representation of a general decomposition of a four-partite state into a mixture of a GME state and biseparable states in the respective bipartition. Here, each vertex corresponds to a party, and connected parties belonging to the same group. (b) A linear minimum covering of a four-partied state, only one link (red, dashed) is cut by the bipartition $\rho_{ABC|D}$. (c) A full covering of a four-partite state, half of the edges are cut by the bipartition $\rho_{ABC|D}$.}
\label{dipict}
\end{figure}

\subsection{DDIC method as a measure of GME}
 For any multipartite quantum state $\rho$, let
\begin{equation}\label{eq: GME+BS}
    \rho = P_{GME} \,\rho_{GME} + P_{BS} \, \rho_{BS}
\end{equation}
with $\rho_{BS}$ biseparable be the decomposition that maximise the biseparable weight $P_{BS} = 1- P_{GME}$. For such state the DDIC score can not exceed $ \bar{\beta}^E \leq P_{GME} \, \beta_Q + P_{BS} \, \bar{\beta}_{BS}^E$ by linearity. Thus the observation of a value for $\bar{\beta}_E$ sets a lower bound
\begin{equation} \label{eq: GME measure}
    P_{GME} \geq \frac{\bar{\beta}^E - \bar{\beta}^E_{BS}}{\beta_Q -\bar{\beta}^E_{BS}}
\end{equation}
on the weight of the GME component in any decomposition of $\rho$ in Eq.~\eqref{eq: GME+BS} (equivalently, an upper-bound on $P_{BS}$). It is easy to see that the minimal weight $P_{GME}$ over all decomposition of a state $\rho$ is a  GME measure~\cite{Ma}: it is by definition nonzero for all GME states and zero for biseparable ones, it is convex, and can not increase under LOCC. Thus, the DDIC method allows us not only to certify GME, but also to quantify it, via Eq.~\eqref{eq: GME measure}. This quantification is both device-independent and scalable with the number of parties $N$.

\subsection{Experimental Results}

\begin{figure}[htbp]
\centering
\includegraphics[width=5in]{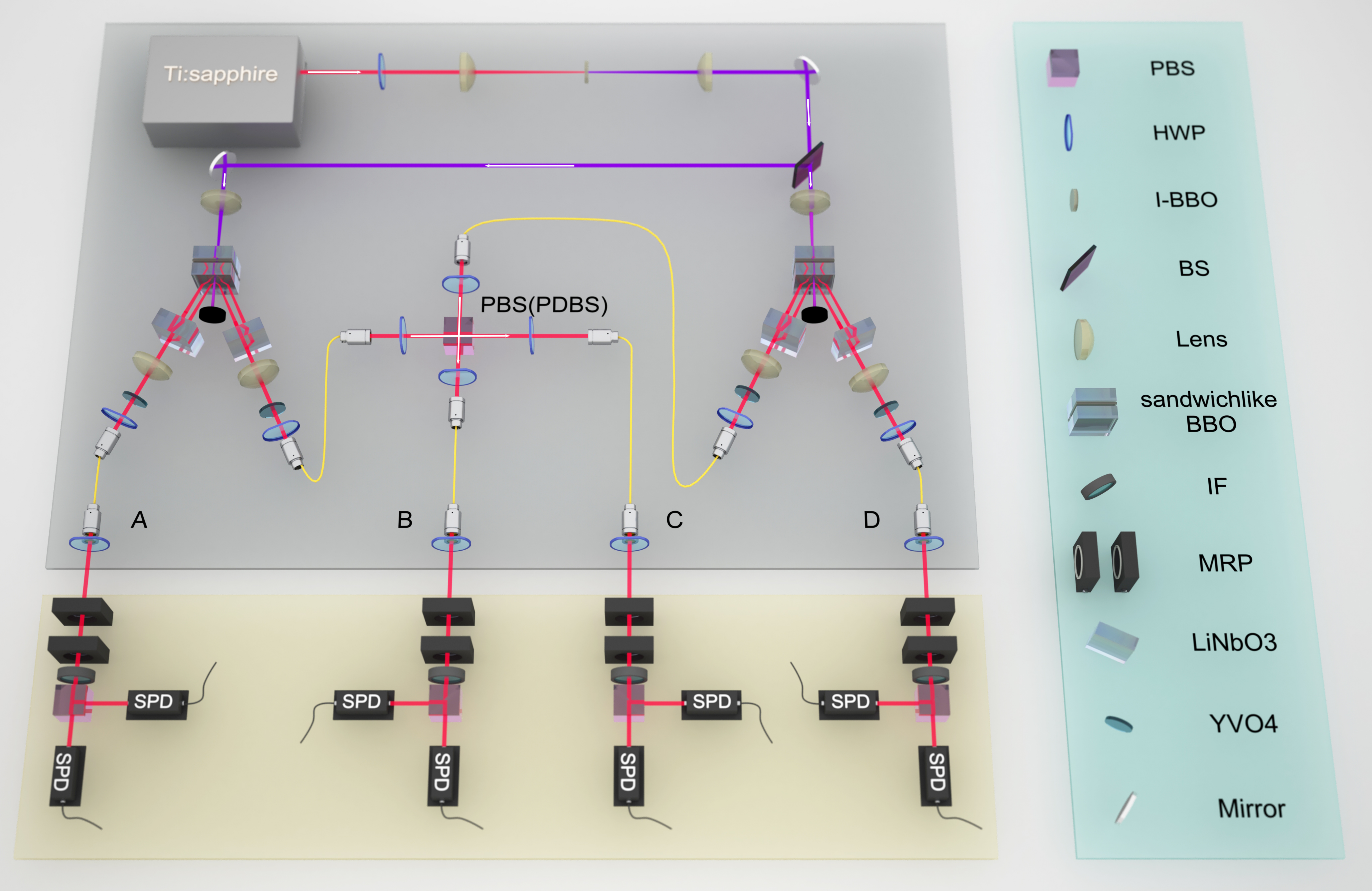}
\caption{\textbf{Experimental setup.} The four-photon states are generated in the grey-shaded part. A frequency-doubled mode-locked Ti:sapphire laser (with a central wavelength of 390 nm, pulse duration of 140 fs and repetition rate of 80 MHz) is split averagely into two beams and used to pump two EPR sources, in each of which a sandwich-like BBO crystal generates polarization-entangled photon pair in state $\alpha|HV\rangle+\beta|VH\rangle$, where H (V) denotes horizontal (vertical) polarization (see Appendix for more details). Various four-photon entangled states are generated by introducing ``interaction" between two uncorrelated photons from two EPR sources. The central polarization beam-splitter (PBS) and polarization-dependent beam-splitter (PDBS) are used to generate generalized GHZ and cluster states respectively. The Bell correlation is measured through four sets of polarization analyzer setup (PAS) shown in the yellow-shaded part. Each PAS consists of elements rotating the polarazion of incident photons in a controlled way, followed by a PBS and two single photon detectors (SPD).}
\label{setup}
\end{figure}

Our basic ingredient to prepare multipartite entangled states is a sandwich-like  EPR source that generates polarization-entangled photon pairs as shown in Fig. ~\ref{setup}. Four-photon entangled states can be prepared by entangling photons from two such independent EPR sources. It is  however{extremely challenging to directly couple individual photons, and the common method is using the measurement induced nonlinearity. In other words we use postselection to project the initial state into the desired entangled state with a certain probability. The projector can be realized by a linear optical network which redistributes the input photons and selects the desired entangled subspace in the output.  In the experiment, we introduce two kinds of optical elements (the PBS and PDBS shown in Fig. \ref{setup}), which can be used to generate different families of four-photon or three-photon entangled states,
including GHZ states, cluster states and  weighted graph states. 
As we will now explain, for each family of states, we test the GME with the DDIC method described above. We emphasise that the post-selection does not compromise the DI analysis for our setup, because the local measurements performed with PAS in Fig.~\ref{setup} satisfy the weak-fair sampling assumption as discussed in methods section. This is however not the case for any measurement.

\begin{figure}[htbp]
\centering
\includegraphics[width=6.5in]{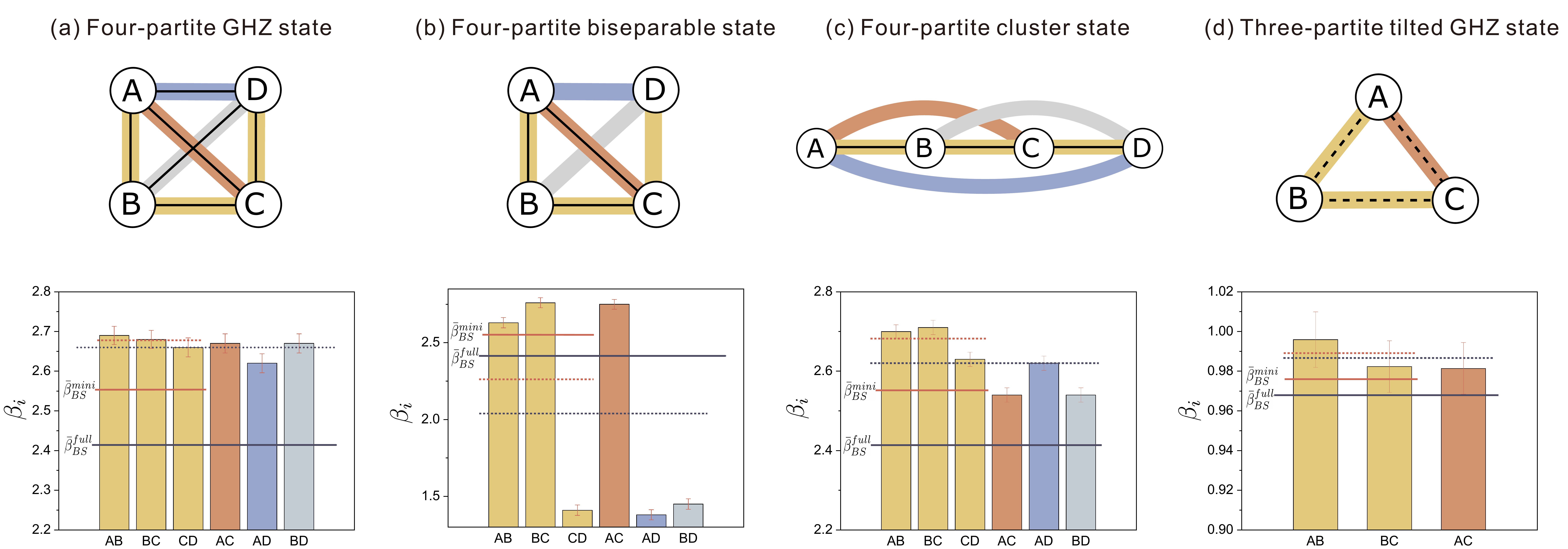}
\caption{ \textbf{DDIC results of several classes of entangled states.} In (a)-(d), pictorial representations of four families of multi-partite entangled states are diagrammed respectively for four-partite GHZ state, four-partite separable state, four-partite cluster state and three-partite tilted GHZ state. Each party is labelled as the circles, the thin solid edges give the graph-state representation of the state, while the thick colored edges give the pairs of parties for the DDIC method. The thick edges in light brown  constitute a minimum covering in which all parties are connected, and all the edges constitute the full covering involving all possible pairs. The histogram below each diagram represents the measured Bell value of each edge in minimum/full coverings. The columns are correspondingly colored to the thick colored edges. The red/blue dotted lines represent the mean Bell values over the pairs in minimum/full coverings, respectively. The red/blue solid lines are the biseparable bounds when measuring the minimum/full coverings, which are obtained from Eqs. \ref{mini} and \ref{full}, respectively. Violation of the biseparable bound by mean Bell value for either minimum/full coverings indicates the presence of GME state.}
\label{struc}
\end{figure}

\emph{Four-photon GHZ state.} The GHZ state $\frac{1}{\sqrt 2} \left( |0\rangle^{\otimes N} +|1\rangle^{\otimes N} \right)$  is a special class of multipartite entangled state, which possesses many particular properties and applications, e.g., GHZ states are the best quantum channels for teleportation \cite{Zhao} and quantum key distribution \cite{Kempe}. Many preparation schemes have been proposed to create GHZ states in different systems, such as cavity QED system \cite{Behzadi}, optical system \cite{Shi} and ion trap system \cite{Monz}. A four-photon GHZ state can be represented as a regular tetrahedron in graph state representation \cite{Hein}, consisting of four vertices (parties) and six edges (pairs). By projecting two parties into the Pauli X basis, the remaining two parties connected by an edge evolve  to maximally entangled two-qubit states in all of the four branches $X^{+}X^{+}$, $X^{+}X^{-}$, $X^{-}X^{+}$, $X^{-}X^{-}$ (where $\pm$ represents the  result of each X-measurement). Interestingly, the states in the 4 branches are related by a Pauli Z transformatin on one of the qubits, and hence can maximally violate the CHSH inequality with the same settings (upon classical relabeling). The CHSH value of each pair $\beta_e$ $(e$=AB,AC,AD,BC,BD and CD) is defined as the averaged violation of CHSH inequality over these  branches. In the experiment, the four-photon GHZ state $|\Psi\rangle_{GHZ}=\frac{1}{\sqrt{2}}(|HHHH\rangle_{ABCD}+|VVVV\rangle_{ABCD})$ is prepared with the fidelity as $\sim0.97$. The CHSH values of all the six pairs are measured and shown in Fig. \ref{struc}(a). To certify GME, the mean CHSH values over the pairs in the minimum and full coverings are calculated to be $\bar{\beta}^E= 2.671\pm0.012$ and $2.662\pm0.009$ respectively, which are plotted as the dotted red/blue lines across the columns forming minimum/full coverings. It can be seen that the measured  average CHSH values  are well above the biseparable bounds for both minimum/full coverings, which are calculated to be 2.552/2.414 shown as the solid red/blue lines. From the full covering we find that the prepared state has $P_{GME}\geq 0.598\pm0.024 $,  after  the filtering defined by post-selection.

\emph{Biseparable state.} To see how to distinguish a biseparable state with  our DDIC method, we prepare a four-photon state $|\Psi\rangle_{sep}=\frac{1}{2}(|HHH\rangle_{ABC}+|VVV\rangle_{ABC})\times
(|H\rangle_{D}+|V\rangle_{D})$, which is a product of a three-photon GHZ state and a single photon state. The DIC is implemented in the same way as the four-photon GHZ state by measuring the mean CHSH values of the minimum/full covering containing three/six pairs of photons, and the measured results of all the six pairs are shown in Fig. \ref{struc}(b). When one pair consists of the isolated photon $D$ and one other photon, it ends up in a separable bipartite state, which is locally bounded and its CHSH value $\beta_{XD}(X=A,B \text{ and } C)$ is below 2. A minimum covering necessarily contains one such separable pair; and thus, the mean CHSH value is calculated as $2.263\pm0.020$ and below the biseparable bound 2.552. Regarding the full covering, three pairs are separable in all the six pairs, and the mean CHSH value is calculated as $2.067\pm0.014$ which is also below the corresponding biseparable bound 2.414. Indeed, no violation is observed across the $ABC|D$ splitting. These results  illustrate how a non-GME state fails to violate the biseparable bound.  On the other hand for the three photon GHZ state prepared on parties ABC we find that $P_{GME}\geq 0.785\pm0.035$.

\emph{Cluster state.} The cluster states, e.g. $|\Psi\rangle_{cluster}=\frac{1}{2}(|0000\rangle_{ABCD}+|0011\rangle_{ABCD}+|1100\rangle_{ABCD}-|1111\rangle_{ABCD})$, has been recognized as
the basic building blocks for one-way quantum computation
which describes a realization of quantum computation beyond
the usual circuit picture \cite{Menicucci1,Menicucci2}. Cluster states are graph states with a lattice graph (with low degree), thus a maximally entangled two-qubit state on an edge $e \in G$ (graph state representation) can be prepared by only measuring the few neighbouring parties. In the case of linear cluster states this requires to measure only one or two parties. Hence for the minimal covering at most four parties have to be measured in each run of the experiment, which is a great asset for scaling up $N$ given that in practice each measurement adds some noise to the state.  In our four-photon experiment,  this can be seen when measuring the pair AB (or CD), where only photon C (or B) need to be measured, and the last photon just acts as a trigger. The produced state is close to the linear cluster state $|\Psi\rangle_{cluster}$ with fidelity $\sim0.95$. The averaged CHSH values for the six pairs are shown in Fig.\ref{struc}(c). The mean CHSH values for the minimum/full coverings are calculated as $2.653\pm0.010$/$2.620\pm0.007$, which violate the biseparable bounds and certify the GME cluster state. The GME weight of the state is found to be $P_{GME}\geq 0.497\pm0.017$.}

\emph{Generalized GHZ state.}
 To illustrate the generality of the DDIC method we now apply it to a weakly entangled three qubit state.  In the experiment, we prepare an extremely tilted GHZ state close to $|\Psi\rangle_{par}=\cos\theta|0\rangle^{\otimes 3}+\sin\theta|1\rangle^{\otimes 3}$ with $\theta=14.3^\circ$. When applying the DDIC method to this state, it is impossible to produce maximally entangled bipartite states on all branches. Instead with a Pauli X measurement on one of the qubits we prepare partially entangled states $\cos(\theta)|00\rangle + \sin(\pm \theta)|11\rangle$. These states violate the CHSH inequality, but not enough for their mean CHSH values to surpass the biseparable bound, even in the ideal case. We thus adjust the DDIC method by replacing the CHSH inequality with the following bipartite inequality, which is maximally violated by partially-entangled states~\cite{1812.02628}:

\begin{equation}
\begin{split}
\label{Itheta}
&\mathcal{I}_{\theta}=\frac{1}{4}\left[\frac{\left\langle A_{0}\left(B_{2}-B_{3}\right)\right\rangle}{\sin \left(b_{\theta}\right)}+\frac{\sin (2 \theta)}{\cos \left(b_{\theta}\right)}\left\langle A_{1}\left(B_{2}+B_{3}\right)\right\rangle+\cos (2 \theta)\left(\left\langle A_{0}\right\rangle+\frac{\left\langle B_{2}-B_{3}\right\rangle}{2 \sin \left(b_{\theta}\right)}\right)\right] \\
&\leq \frac{1}{4}\left[\cos (2 \theta)+(2+\cos (2 \theta)) \sqrt{\frac{7-\cos (4 \theta)}{5+\cos (4 \theta)}}\right], \\
\end{split}
\end{equation}
with $b_{\theta}=\arctan \sqrt{\left(1+\frac{1}{2} \cos ^{2}(2 \theta)\right) / \sin ^{2}(2 \theta)}$.
The prepared state is tested with the inequality $\mathcal{I}_{15^\circ}$ of which we have $\beta^Q=1$ and $\beta^L=0.952$, and the biseparable bounds of minimum and full coverings are determined as 0.976 and 0.968 respectively.
The experimental results shown in Fig. \ref{struc}(d) certify the genuine entanglement in the partially entangled GHZ state, since the biseparable bounds are distinctly violated by the mean Bell values, which are calculated as $0.989\pm0.010$/$0.987\pm0.008$ of minimum/full coverings. Quantitatively, we find that $P_{GME} \geq 0.594\pm0.250$.

Note that the considered weakly entangled GHZ state with $\theta=14.3^\circ$ cannot violate the standard Svetlichny inequality~\cite{Svetlichny}, because a Svetlichny model can reproduce all tripartite correlations for this state when $\theta$ is below $15^\circ$~\cite{Zukowski}. A dedicated Svetlichny-type inequality would thus be needed to demonstrate genuine nonlocality, such as the one given in~\cite{Bancal10}. Here, we are able to demonstrate genuine entanglement by simply choosing adequate bipartite Bell inequality, hence demonstrating the flexibility of our method for the detection and quantification of GME states.

\section{Discussion}

The DDIC method provides a way for reliable GME certification in a wide range of states. It also enables the quantification of genuine multipartite entanglement via the weight of the minimal GME component, and moreover is both device-independent and intrinsically resistant to realistic noise. This allows us to demonstrate and quantify GME experimentally in a variety of multipartite states, including in a genuinely but weakly entangled state.

The DDIC method infers properties of multipartites states by leveraging bipartite Bell tests in an optimal way. When applied to generalized and weighted graph states with bounded degree, this allows the method to involve at most a constant number of parties in each run of the experiment. At the same time, this fundamentally limits the way that noise resistance scales with the number of parties. It would be interesting to see if this limitation can be overcome by a DDIC construction going beyond bipartite primitives. Can such a method achieve a constant noise tolerance? Given the wide applicability of the DDIC method to various states, and thus its unique relevance to quantum science, it would also be fruitful to explore the relation between the DDIC method and self-testing.






\section{Acknowledgments}

This work was supported by the National Key Research and Development Program of China (Nos. 2016YFA0302700, 2017YFA0304100), National Natural Science Foundation of China (Grant Nos. 11874344, 61835004, 61327901, 11774335, 91536219, 11821404), Key Research Program of Frontier Sciences, CAS (No. QYZDY-SSW-SLH003), Anhui Initiative in Quantum Information Technologies (AHY020100, AHY060300), the Fundamental Research Funds for the Central Universities (Grant No. WK2030020019, WK2470000026), Science Foundation of the CAS (No. ZDRW-XH-2019-1). W.-H.Z. and Z.C. contribute equally to this work.

\section{Methods}

\subsection{Weak fair-sampling assumption and post-selection.}

The DI analysis of an experiment requires at least some parties to chose the measurement settings randomly, i.e. independently of each other and of the state preparation. Otherwise, the preparation and the measurement devices could easily conspire and fake any quantum correlation with pre-established agreements. For the same reason in the DI analysis of an experiment one can not reject measurement events without compromising the conclusions, an observation known as detection loophole. Indeed, if some measurement outcomes are simply ignored (e.g. no-click events where no photons are detected by some party) it could be that the detector only gives a meaningful outcome upon receiving the desired measurement setting, which effectively compromises the independence of the measurement settings. Yet, in practice photon detectors have a limited efficiency. Hence, no-click events are unavoidable and their rate increases with the number of parties in the measured state. Even in the bipartite case this makes experiments showing loophole free violation of Bell tests extremely challenging.

A way to circumvent this difficulty is to introduce some minimal well-justified assumptions on the internal function of the detectors, which greatly simplifies the detection of GME states in practice but keeps some of the robusteness proper to the fully DI approach. In particular, this can be achieved with the help of the fair-sampling type assumptions \cite{Berry,Orsucci}. 
A measurement device satisfies the weak fair-sampling assumption~\cite{Orsucci}, if the process responsible for the occurrence of no-click events is independent of the choice of the measurement setting. For our purpose this can be formalized in two equivalent ways. First, any measurement $\mathcal{M}$ in the quantum formalism admits a POVM model $\{E_{a|x}\}_{a,x}$, where $x$ labels the measurement setting, $a$ the measurement outcome, and the positive Hermitian operators $E_{a|x}$ satisfy $\sum_a E_{a|x}=1$. In our case, for all measurement settings one of the outcomes $a=\emptyset$ corresponds to a no-click event. The measurement $\mathcal{M}$ is said to satisfy the weak fair-sampling assumption if the POVM element corresponding to all no-click event is the same for all measurement setting
\begin{equation} \label{eq: fair-sampling}
E_{\emptyset|x}= E_{\emptyset|x'} \qquad \forall \, x,x'.
\end{equation}
Equivalently, any such measurement $\mathcal{M}= \overline{\mathcal{M}} \circ F$ can be decomposed as a probabilistic filter $F$ (which can output a no-click outcome and is the same for all measurement settings) followed by a ideal measurement $\overline{\mathcal{M}}$ (which does not have a no-click outcome).

If all the measurement devices in the experiment satisfy this assumption, it can be shown (see \cite{Orsucci}), that the post-selected statistics on the produced state $\rho$ is identical to the statistics produced by the filtered state $\rho^F = \frac{(F^{(1)}\otimes\dots\otimes F^{(N)})[\rho]}{\textrm{tr} (F^{(1)}\otimes\dots\otimes F^{(N)})[\rho]}$ with the some ideal measurements (which always click and do not open the detection loophole).  Thus our method applied to the post-selected statistics reveals that the filtered state $\rho^F$ is GME. But local filtering can not create entanglement, and  the original state $\rho$ prepared in the experiment is also GME.  This shows that when the measurement apparatus satisfies the weak fair sampling assumption one can reject the no-click events from the analyzed measurement data without compromising the GME proof.  On the other hand the amount of GME, as quantified by the GME-weight in Eq.~\eqref{eq: GME measure}, relates to the filtered state $\rho^F$.

Note that the weak fair-sampling assumption only requires the probability of no-click to be independent of the measurement setting (for all possible input states) and does not require a precise model or calibration of the measurement devices. In our experiment with photons entangled in polarization, each local PAS measurement consist of two non-number-resolving single photon detector preceded by a polarizing beamsplitter. The measurement setting is set by a elements rotating the polarization of the incoming photons before the PBS. Very generally, one models the no-detection outcomes as events where all photons have been lost. Since the transmission of the element rotating the polarization is constant, and the efficiency of the two detectors are the same~\footnote{We note that one can also make quantitative statements in the case where the fair-sampling assumption is only verified approximately, for example the efficiency of the two detectors are approximately equal or the transmission can only depend mildly on the setting.} the occurrence of the no-click events (photon loss) is manifestly independent of the measurement setting, see \cite{Orsucci} for the formal argument. Hence, the weak fair-sampling assumption is well justified in our setup.

\section{Appendix}
\label{app: FS}

\subsection{Optimality of the biseparable bound for the full coverings}
Let us show now that that the biseparable bound
\begin{equation}
    \bar{\beta}^{full}_{BS} = \beta^Q - \frac{2}{N}(\beta^Q -\beta^L)
\end{equation}
found for the full coverings is optimal. That is, there is no covering $E$ for which the minimal average violation certifying GME is lower. To do so, we first note that the biseparable bound for a general covering $E$ satisfies
\begin{equation}
    \bar{\beta}^{E}_{BS} = \frac{(|E|-\text{mincut}[E])\beta^Q + \text{mincut}[E] \beta^L}{|E|} = \beta^Q -\frac{\text{mincut}[E]}{|E|} (\beta^Q - \beta^L),
\end{equation}
where $\text{mincut}[E]$ is the minimal number of edges cut by a bipartitition of the parties in two disjoint groups $g_1|g_2$. Denoting by $\text{neighbourhood}[i]$ the number of edges in $E$ involving the party $i$, one easily sees that
\begin{equation}
    \text{mincut}[E] \leq \min_{i=1,\dots N} \text{neighbourhood}[i],
\end{equation}
as $\{i\}| \{1,\dots, N\}\setminus i$ are valid  bipartitions. Now, the total number of edges in $E$ satisfies
\begin{equation}
    |E| = \frac{1}{2} \sum_{i=1}^N \text{neighbourhood}[i] \geq \frac{N}{2} \left(\min_{i=1,\dots N} \text{neighbourhood}[i] \right).
\end{equation}
Finally, using the last two inequalities one gets
\begin{equation}
    \frac{\text{mincut}[E]}{|E|} \leq  \frac{\min_{i=1,\dots N} \, \text{neighbourhood}[i]}{\frac{N}{2} \min_{i=1,\dots N} \text{neighbourhood}[i]} = \frac{2}{N}.
\end{equation}
For the biseparable bound this implies
\begin{equation}
    \bar{\beta}^{E}_{BS}  \geq \beta^Q - \frac{2}{N} (\beta^Q-\beta^L) = \bar{\beta}^{full}_{BS}
\end{equation}
and concludes the proof.

\subsection{The ring covering}
The biseparable bound $\bar{\beta}^{E}_{BS}$ on the average violation for the DDIC method depends on the choice of the covering $E$. In the main text we have presented the two extreme cases such that $\bar{\beta}^{mini}_{BS}\geq \bar{\beta}^{E}_{BS} \geq \bar{\beta}^{full}_{BS}$. The minimal covering involves the minimal number of edges $|E_{mini}|=N-1$ required to conclude about GME, but leads to the highers biseparable bound~\eqref{mini}. The full covering requires to measure all pairs of parties $|E_{full}|=\frac{1}{2}N(N-1)$, but leads to the lowest possible biseparable bound~\eqref{full}. In general, the choice of the covering has to be adjusted to the experiment in consideration, given that some bipartite entangled states are harder to prepare and suffer more from noise depending on the structure of the state.
\begin{figure}
    \centering
    \includegraphics{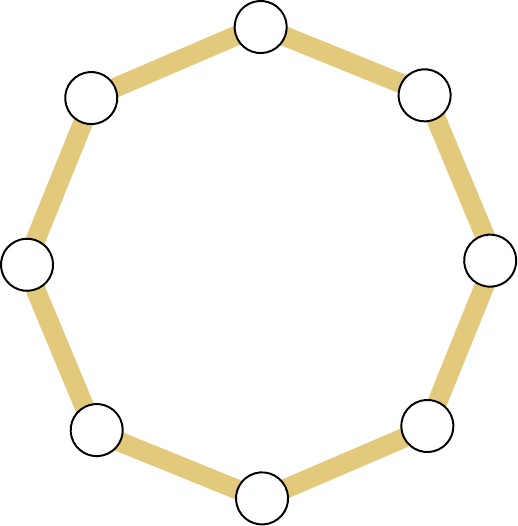}
    \caption{The ring covering $E_{ring}$ for $N=8$.}
    \label{fig:ring}
\end{figure}
Here we present the example of a ring covering $E_{ring}$, sketched in Fig.~\ref{fig:ring}. For $N$ parties labeled by $i =1,\dots N$, the ring covering is defined by a 1-D lattice with periodic boundary condition (a ring), made of nearest-neighbour edges $\{i, i+1\,  \text{mod}\,  N\}\in E_{ring}$. The size of the covering $|E_{ring}| = N = |E_{mini}|+1$ grows linearly with $N$ similarly to $E_{mini}$. Furthermore, for $N\geq 3$ it is easy to see that any bipartition $g_1|g_2$ cuts at least two edges, and the biseparable bound reads
\begin{equation}
\bar{\beta}_{BS}^{ring} = \frac{(|E_{ring}|-2)\beta^Q + 2 \beta^L}{|E_{ring}|} =\beta^Q - 2 \frac{\beta^Q-\beta^L}{N} = \bar{\beta}^{full}_{BS}.
\end{equation}
Manifestly, the ring covering gives the same optimal biseparable bound than the full covering, but only requires to measure $N$ edges, combining the advantages of $E_{full}$ and $E_{mini}$.

\subsection{Experimental details}

{\it EPR source.---}We employ a sandwich-like EPR source in the experiment, the detailed structure is shown in Fig.S1. A frequency-doubled mode-locked Ti:sapphire laser (with a central wavelength of 390 nm, pulse duration of 140 fs and repetition rate of 80 MHz) is focused on the sandwich-like crystals to produce photon pairs. The BBO1 and BBO2 are identically cut for beamlike type-II phase-matching. The pump photon is equal probability to be downconverted in the two BBO crystals which both produce polarization states $|H\rangle_{e,2}|V\rangle_{o,1}$, the subscript o (e) denotes the ordinary (extraordinary) photon with respect to the BBO crystal. A true-zero-order half-wave plate (THWP) is inserted in the middle to rotate the polarization state of the photon pairs produced by the first BBO to $|H\rangle_{e,2}|V\rangle_{o,1}$. After spatial and temporal compensations, the produced polarization state becomes entangled state $\alpha|H\rangle_{e,2}|V\rangle_{o,1}+\beta|V\rangle_{e,2}|H\rangle_{o,1}$. The sandwich structure engineers the e- and o-polarized photons into different spatial modes, which meets the key requirement of the entanglement concentration scheme and make a perfect polarization entanglement in principle achievable.
In the experiment, in order to tune the ratio $|\alpha|^2/|\beta|^2$, we use two BBO crystals of different thickness. The detailed thickness of the BBO crystals and the corresponding compensations are listed in Table.S1.

\begin{table}[htb]
\centering
\caption{The detailed thickness of crystals for producing different two-photon states. LiNbO3 (YVO4) crystals are used for spatial (temporal) compensations. e (o) represents the compensation crystal for extraordinary (ordinary) photon. Note that for thicker BBO crystals, the spectrum width of down-converted photons will become narrower, while we use fixed narrow-band filters, thus the photon pair generation rate will not increase linearly with the crystal thickness.}
\begin{tabular}{|c|c|c|c|c|c|c|} \hline
State & BBO1 & BBO2 & LiNbO3 (e) & LiNbO3 (o) & YVO4 (e) & YVO4 (o) \\\hline
$\frac{1}{\sqrt{2}}|HH\rangle+\frac{1}{\sqrt{2}}|VV\rangle$ & 1 mm & 1 mm & 3.2 mm & 1 mm & 0.42 mm & 0.6 mm \\\hline
$\frac{1}{2}|HH\rangle+\frac{\sqrt{3}}{2}|VV\rangle$ & 2 mm & 1 mm & 4.2 mm & 0.5 mm & 0.47 mm & 0.57 mm \\\hline
\end{tabular}
\end{table}

\begin{figure}[b]
\centering
\includegraphics[width=0.5\textwidth]{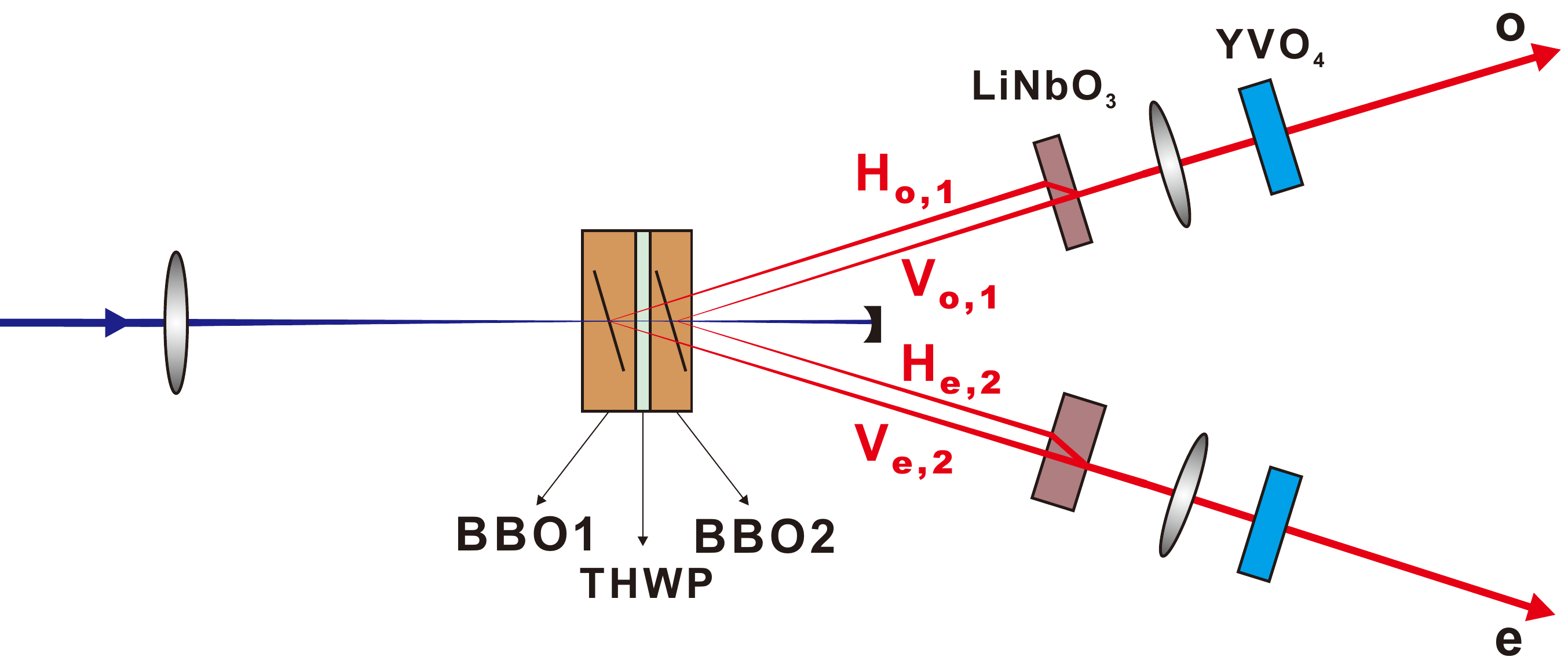}
\caption{\label{Fig:S1} EPR source.}  \end{figure}

{\it Generation multiphoton entangled states.---}As shown in Fig.2 of the main text, we introduce two kinds of optical elements to connect two EPR sources to generate different kinds of multiphoton entangled states.

We use polarization beam-splitter (PBS) to generate the four-photon GHZ state and the three-photon tilted GHZ state. The input state is $\frac{1}{\sqrt{2}}(|HH\rangle+|VV\rangle)_{12}\otimes\frac{1}{\sqrt{2}}(|HH\rangle+|VV\rangle)_{34}$ or $\frac{1}{\sqrt{2}}(|HH\rangle+|VV\rangle)_{12}\otimes(\alpha|H\rangle+\beta|V\rangle)_{34}$ .  The PBS is a polarization component which transmits the H polarizations while reflects the V polarizations. Only the two input photons have the same polarization can they be transmitted or reflected by the PBS and lead to a coincidence at each output. Thus by postselecting there is one and only one photon in each output port, the PBS acts as a parity check operator $|HH
\rangle\langle HH|+|VV\rangle\langle VV|$. It is easy to check that by performing a parity check operator between qubit 2 and 3, the target states can be prepared.

We use polarization dependent beam-splitter (PDBS) to produce the four-photon linear cluster state. Such a component is designed to realize an optical Control-phase gate. Its transmission efficiencies for H and V polarizations are set to $T_H=1$ and $T_V=1/3$ respectively. Thus when the input two-photon state is $|VV\rangle$, the PDBS acts as a partial beam-splitter which will introduce a $\pi$ phase shift due to the HOM interference, while for the input state $|HH\rangle$, $|HV\rangle$ or $|VH\rangle$, there is no interference and the PDBS only attenuates the V polarized components. Such a performance is just like a C-phase gate. In order to equally attenuate the H polarized components, two additional PDBSs with complementary transmissions $T_H=1/3, T_V=1$ are placed at the two output ports. The scheme to generate a four-photon linear cluster state is straight forward---by performing a C-phase gate between two Bell states. In the experiment, we further integrate the two additional PDBSs into the sources. Thus the input state becomes $|\psi\rangle_{in}=\left(\frac{1}{2}|HH\rangle+\frac{\sqrt{3}}{2}|VV\rangle\right)_{12}\otimes\left(\frac{1}{2}|HH\rangle+\frac{\sqrt{3}}{2}|VV\rangle\right)_{34}$,
Then the neighboring photons 2 and 3 are overlapped on the PDBS for interference.
The mode transformation of the PDBS is
\begin{equation}
\left\{
\begin{aligned}
\label{eq:PDBS}
h^\dag_2 &\rightarrow h^\dag_{2'}\\
h^\dag_3 &\rightarrow h^\dag_{3'}\\
v^\dag_2 &\rightarrow \sqrt{\frac{1}{3}} v^\dag_{2'} + i\sqrt{\frac{2}{3}}v^\dag_{3'}\\
v^\dag_3 &\rightarrow \sqrt{\frac{1}{3}} v^\dag_{3'} + i\sqrt{\frac{2}{3}}v^\dag_{2'}
\end{aligned}
\right.
\end{equation}
where $h^\dag_{i(i')}$ and $v^\dag_{i(i')}$ denote the creation operators for H and V polarized photons in the $i(i')$ spatial mode.
One can check that by postselecting there is one and only one photon in each output port, the state can be projected into the linear cluster state $|\psi\rangle_{out}=\frac{1}{2}(|HHHH\rangle+|HHVV\rangle+|VVHH\rangle-|VVVV\rangle)$. By removing the attenuation of the H polarizations and using nonmaximally entangled states as input, the projective probability dramatically increased from 1/9 to 1/4.


{\it Detection.---} The qubit analyser consists of a QWP, a HWP, a PBS and two fiber-coupled single photon detectors. Such a device can project the single-qubit polarization state onto any desired direction on the Bloch sphere.

{\it HOM interference.---}To generate entanglement between independent photon sources, there is always HOM interference in the optical network which requires the indistinguishability of the ``interacting" photons. In the experiment, we use 1-nm (3-nm) bandwidths filters for the interfering (trigger) photons for spectral selection. With
such settings, we observe the HOM interference with a visibility of $97.9\%$ (see Fig.S2).

\begin{figure}[tb]
\centering
\includegraphics[width=0.5\textwidth]{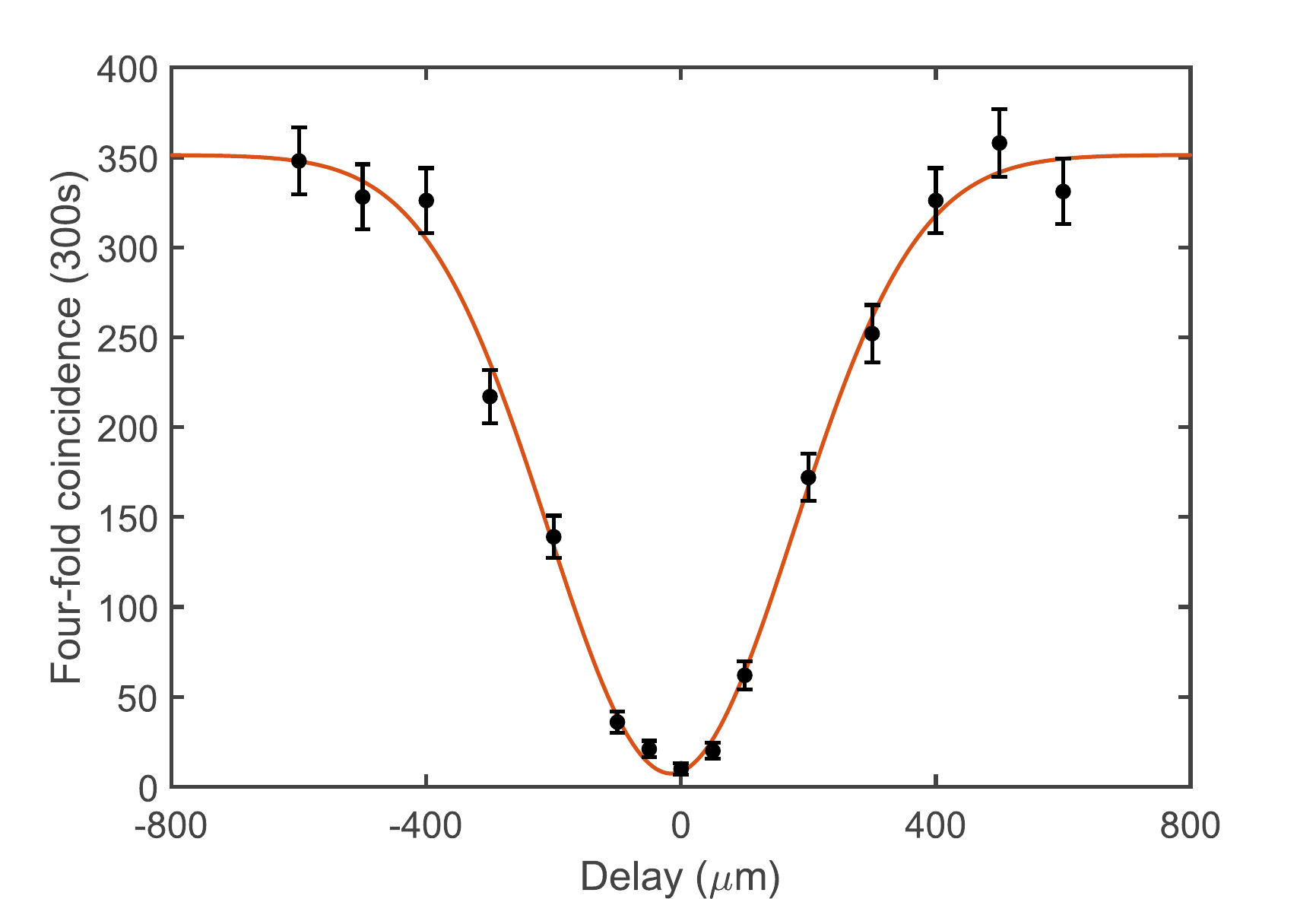}
\caption{\label{Fig:S1}  HOM interference fringe.}  \end{figure}

{\it Characterize the prepared entangled states.---}Significant effort is made to prepare high fidelity entangled states, especially the tilted GHZ state, which has a weak violation of the Bell inequalities. In the experiment, we use a pump power of 50 mW for each source, the counting rate of each source is about 10 kHz. Fig.S3 shows the tomographic results of all the tested states in the experiment.

\begin{figure}[tb]
\centering
\includegraphics[width=0.9\textwidth]{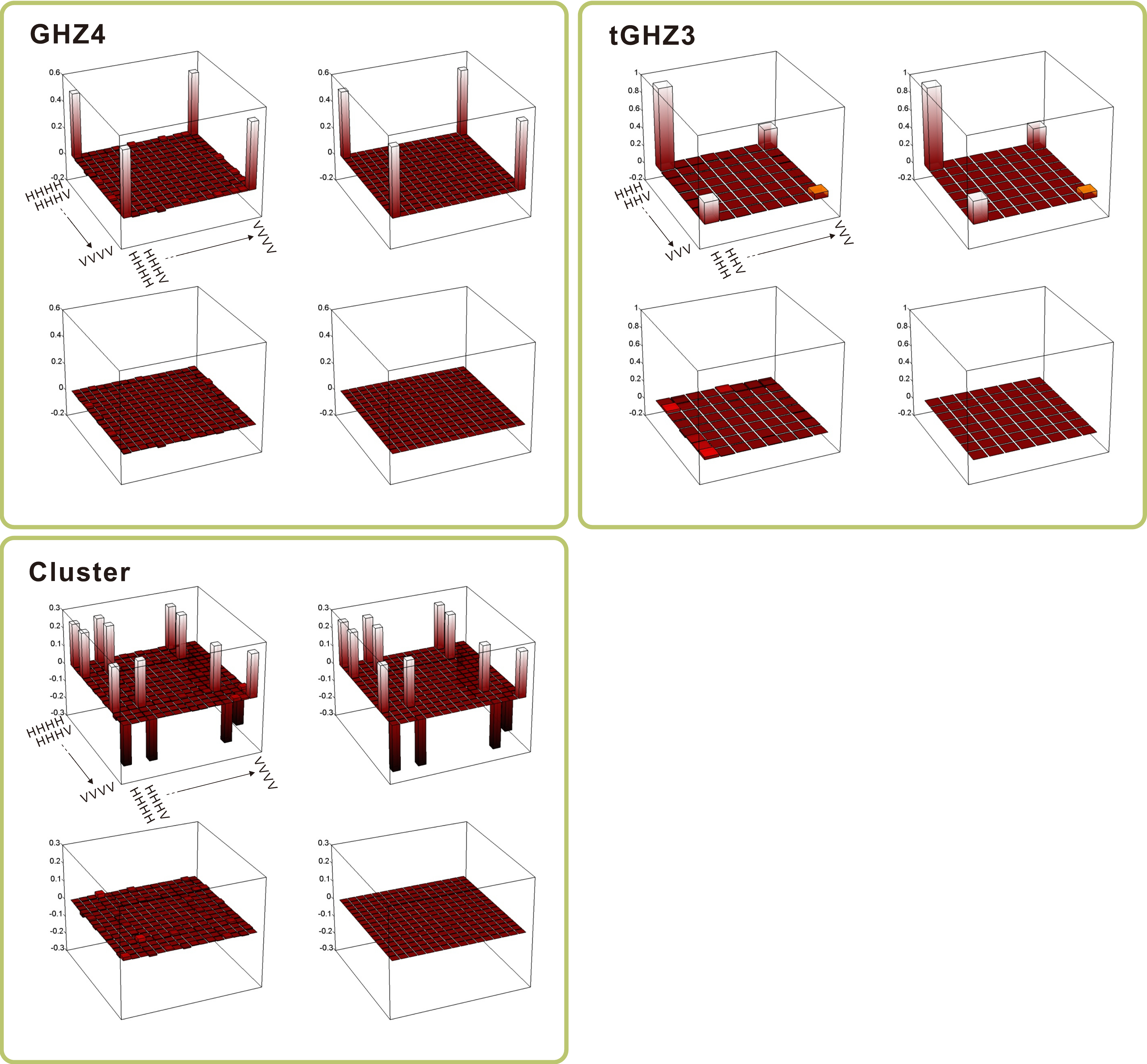}
\caption{\label{Fig:S1}  Tomographic results for the four-photon GHZ state, three-photon tilted-GHZ state and four-photon linear cluster state. In each box, the left two pictures show the real (top) and imaginary (bottom) part
of the experimental reconstructed density matrix, the right pictures show the theoretical density matrix. The fidelity of the
states are $F_{GHZ4}=0.9716\pm0.0030, F_{tGHZ3}=0.9864\pm0.0040, F_{cluster}=0.9437\pm0.0042$ respectively.}  \end{figure}

\subsection{Detailed measurement strategy for each states}

{\it General GHZ states.---} To measure the bipartite Bell violation $\beta_i$ for each edge of the N-qubit GHZ state, all the remaining N-2 parties need to be measured in the Pauli X basis. The two outcomes are labeled by $+1$ and $-1$ for each party. According to the product of the outcomes, the bipartite state becomes either $\frac{1}{\sqrt{2}}(|00\rangle+|11\rangle)$ or $\frac{1}{\sqrt{2}}(|00\rangle-|11\rangle)$, thus we can categorize the results into only two branches after the LOCC protocol. The CHSH operators for the two states are $S_+=A_0B_0+A_0B_1+A_1B_0-A_1B_1$ and $S_-=A_0B_0+A_0B_1-A_1B_0+A_1B_1$ respectively, where $A_0=\sigma_z$, $A_1=\sigma_x$, $B_0=\frac{1}{\sqrt{2}}(\sigma_z+\sigma_x)$, $B_1=\frac{1}{\sqrt{2}}(\sigma_z-\sigma_x)$. Thus we can use the same measurement settings to measure the Bell violation for each edge. It is easy to see the measurement setting is linear scaling $4(N-1)$ for minimal covering.

The analysis is similar for generalized GHZ state. In the experiment we generate an extremely tilted GHZ state $|\Psi\rangle_{par}=\cos\theta|0\rangle^{\otimes 3}+\sin\theta|1\rangle^{\otimes 3}$ with $\theta=15^\circ$. After single-qubit projection, the produced bipartite state is very partially-entangled. If we still use CHSH inequality the biseparable bound cannot be violated even in the ideal case, while the tilted CHSH inequality is very sensitive to noise. We circumvent this problem by using an alternative Bell inequality which is maximally violated by partially-entangled two-qubit state (Eq.8 of the main text). The Bell expression has a quantum bound of 1, achieved by the observables $A_0=\sigma_z$, $A_1=\sigma_x$, $B_0=\cos(b_\theta)\sigma_x+\sin(b_\theta)\sigma_z$, $B_1=\cos(b_\theta)\sigma_x-\sin(b_\theta)\sigma_z$.

{\it Cluster state.---} The measurement setting for N-qubit cluster state is also $4(N-1)$ for minimal covering. After local Pauli measurements on the remaining parties, the bipartite state of the edge is equal to a local unitary applied on the two-qubit cluster state, i.e., either $\frac{1}{\sqrt{2}}(|0+\rangle\pm|1-\rangle)$ or $\frac{1}{\sqrt{2}}(|0-\rangle\pm|1+\rangle)$. The Bell violation for each branch can also be measured by using the same four measurement settings. Also, only the neighboring parties need to be measured in the LOCC protocol for each pair. This is due to the cluster will be decoupled after Pauli Z measurement on the site.

{}

\end{document}